\documentclass[conference]{IEEEtran}
\IEEEoverridecommandlockouts

\usepackage{graphicx}
\usepackage{amsmath, amsfonts}
\usepackage{amsthm}
\usepackage{mathtools}
\usepackage{amssymb}
\usepackage{lscape}
\usepackage{verbatim}
\usepackage[caption=false,font=normalsize,labelfont=sf,textfont=sf]{subfig}
\usepackage{multirow}
\usepackage{algpseudocode}
\usepackage[linesnumbered,ruled,vlined]{algorithm2e}
\usepackage{braket}

\usepackage{array}
\usepackage{textcomp}
\usepackage{stfloats}
\usepackage{url}
\usepackage{balance}
\usepackage{cite}

\newtheorem{myDef}{Definition}

\newtheorem{myProposition}{Proposition}

\usepackage{xcolor}
\def\BibTeX{{\rm B\kern-.05em{\sc i\kern-.025em b}\kern-.08em
    T\kern-.1667em\lower.7ex\hbox{E}\kern-.125emX}}
\begin{document}

\title{
Construction A Lattice Design\\ Based on the Truncated Union Bound \\
\thanks{This work was supported by JSPS Kakenhi Grant Number JP 21H04873.}
}

\author{\IEEEauthorblockN{Jiajie Xue and Brian M.~Kurkoski}
\IEEEauthorblockA{\textit{School of Information Science} \\
\textit{Japan Advanced Institute of Science and Technology}\\
Nomi, Japan \\
\{xue.jiajie, kurkoski\}$@$jaist.ac.jp}
\and
\IEEEauthorblockN{Emanuele Viterbo}
\IEEEauthorblockA{\textit{ECSE Dept.} \\
\textit{Monash University}\\
Clayton, Australia \\
{emanuele.viterbo$@$monash.edu} }
}

\maketitle

\begin{abstract}
This paper considers $n= 128$ dimensional construction A lattice design, using binary codes with known minimum Hamming distance and codeword multiplicity, the number of minimum weight codeword. 
A truncated theta series of the lattice is explicitly given to obtain the truncated union bound to estimate the word error rate under maximum likelihood decoding. The best component code is selected by minimizing the required volume-to-noise ratio (VNR) for a target word error rate $P_e$. The estimate becomes accurate for $P_e \leq 10^{-4}$, and design examples are given with the best extended BCH codes and polar codes for $P_e= 10^{-4}$ to $10^{-8}$. A lower error rate is achieved compared to that by the classic balanced distance rule and the equal error probability rule. The $(128, 106, 8)$ EBCH code gives the best-known $n=128$ construction A lattice at $P_e= 10^{-5}$.

\end{abstract}


\section{Introduction}
A lattice $\Lambda$ is a discrete additive subgroup of the Euclidean space $\mathbb{R}^N$. 
For power constrained communications, a lattice code $\Lambda_c/ \Lambda_s$ with finite constellation can be constructed by a coding lattice $\Lambda_c$ and a shaping lattice $\Lambda_s$. Lattice codes can achieve the additive white Gaussian noise (AWGN) channel capacity for asymptotically large dimension \cite{erez2004achieving} and have found novel applications in network coding such as compute-forward relaying \cite{nazer2011compute}.

The error performance of a lattice code depends on the underlying coding lattice, which is the design target of this paper.
For dimension $n \leq 24$, Conway and Sloane summarized the best-known lattices in \cite{conway1993sphere} with well-studied properties, such as their shortest lattice vectors. 
For medium to large dimensions, construction A \cite{conway1993sphere} and construction D/D' \cite{barnes1983new} form lattices using a linear block code and a family of nested binary codes, respectively. 
For dimension $n \geq 1000$, LDA lattices \cite{di2012integer} and polar code lattices \cite{liu2018construction} are good lattices for communications. 

This paper focuses on medium dimension using $n= 128$, which allows us to give more precise design and reduces the decoding latency compared to $n> 1000$. Previous studies in this dimension considered construction D lattices using extended BCH (EBCH) codes \cite{matsumine2018construction} and polar codes \cite{ludwiniananda2021design}. However, there has been relatively little study for construction A, which can achieve lower decoding complexity than construction D/D' by only using one component code. Additionally, construction A lattices can be simultaneously good as sphere packing, covering, channel coding and mean-squared error quantization for asymptotically large dimension in \cite{erez2005lattices}, which motivated us to study the performance in finite dimension.

This paper considers finite dimensional construction A lattice design and aims to give the best-known construction A lattice in dimension $n= 128$. In classic design, the balanced distance rule and the equal error probability rule are widely used \cite{wachsmann1999multilevel}. In this paper, the truncated union bound-based design is considered so that lattices can achieve a near-optimal error performance under the maximum likelihood (ML) decoding. Binary linear block codes are applied as the component codes, widely used in practical applications and have codeword Hamming weight equal to its Euclidean norm. In particular, the design considers binary codes with known minimum Hamming distance $d_c$ and the codeword multiplicity $\tau_c$, that is the number of codeword at minimum Hamming weight $d_c$. A truncated theta series of lattices, that is the weight enumerator function of lattices in the Euclidean space, is explicitly given for terms with Euclidean norm $d^2 \leq d_c$. 
From that, the truncated union bound is computed to estimate the word error rate (WER) under ML decoding. The best lattice is found by minimizing the required volume-to-noise ratio (VNR) for a target error rate. Design examples are given using EBCH codes and polar codes as component codes, since EBCH codes have a large minimum Hamming distance and polar codes \cite{arikan2009channel} are accepted as a part of the channel coding scheme in the 5G standard. The codeword multiplicity $\tau_c$ of $n= 128$ EBCH codes are summarized in \cite{BCHweight2022}. A set of polar codes is designed in \cite{rowshan2023formation} based on the partial orders of the binary representation of bit-channel indices, for which $\tau_c$ can be computed analytically. 

Numerical evaluations apply the ordered-statistics decoding (OSD) algorithm \cite{fossorier1995soft} to decode binary codes to give a near-ML performance. 
Since the truncated union bound gives a good estimate for WER of $P_e \leq 10^{-4}$, the design examples are given for target $P_e= 10^{-4}$ to $10^{-8}$.
The best construction A lattices at different target $P_e$ are found, for which each lattice has the lowest required VNR to achieve a desired $P_e$, using EBCH codes and polar codes, respectively. 
For design using EBCH codes, the $(128, 106, 8)$ EBCH code lattice has the lowest required VNR at $P_e= 10^{-4}$ to $10^{-6}$; and the $(128, 113, 6)$ EBCH code lattice has the lowest required VNR at $P_e= 10^{-7}$ to $10^{-8}$. For design using polar codes, the lowest required VNR is obtained by the $(128, 99, 8)$ polar code lattice at $P_e= 10^{-4}$ to $10^{-8}$. 
It is also shown that construction A lattices based on the proposed method achieve lower WER than those based on the classic balanced distance rule and the equal error probability rule, using EBCH codes and polar codes. 
At $P_e= 10^{-5}$, the $(128, 106, 8)$ EBCH code gives the best-known construction A lattice in $n= 128$, which has the lowest required VNR, among our design examples using EBCH codes and polar codes.


The organization of this paper is as follows. Section~\ref{sec_pre} introduces definitions of lattices and construction A. Section~\ref{sec_design_method} gives the design method using the truncated union bound estimate. Section~\ref{sec_design_exampla} gives design examples using EBCH codes and polar codes. Section~\ref{sec_sim} compares the WER performance using different design rules.




\section{Preliminaries} \label{sec_pre}
This section gives a review of definitions of lattices, construction A and its encoding/decoding algorithms.
\subsection{Lattices}
\begin{myDef} \label{def_lattice}
    \rm (Lattice) A lattice is a discrete additive subgroup of the real number space $\mathbb{R}^n$. Let $\mathbf{g}_1, \mathbf{g}_2, \cdots, \mathbf{g}_n \in \mathbb{R}^n$ be $n$ linearly independent column vectors and a generator matrix $\mathbf{G}= [\mathbf{g}_1, \mathbf{g}_2, \cdots, \mathbf{g}_n]$. A lattice $\Lambda$ is defined as: 
\begin{align}
    \Lambda= \Set{\mathbf{Gb}| \mathbf{b} \in \mathbb{Z}^n}.
\end{align}
\end{myDef} 
For each lattice point $\mathbf{x} \in \Lambda$, the Voronoi region of $\mathbf{x}$ consists of $\mathbf{y} \in \mathbb{R}^n$ closer to $\mathbf{x}$ than any other lattice points $\mathbf{x}' \neq \mathbf{x}$, defined as
\begin{align}
    \mathcal{V}(\mathbf{x})= \Set{\mathbf{y} \in \mathbb{R}^N | \|\mathbf{x}- \mathbf{y}\|^2 \leq \|\mathbf{x}'- \mathbf{y}\|^2, \mathbf{x}' \neq \mathbf{x}}.
\end{align}
The Voronoi regions of different lattice points do not overlap and have the same volume given as: 
\begin{align}
    V(\mathcal{V}(\mathbf{x}))= |\det(\mathbf{G})|,
\end{align}
which is also referred to as the volume of the lattice $\Lambda$, denoted as $V(\Lambda)$. A lattice has an infinite constellation thus is power unconstrained. To evaluate the error performance of a lattice, the volume-to-noise ratio (VNR) is defined as:
\begin{align} \label{equ_VNR}
    \text{VNR}= \frac{V(\Lambda)^{(2/ n)}}{2\pi e \sigma^2},
\end{align}
where $\sigma^2$ is the per-dimensional Gaussian noise variance. 

The theta series is the weight enumerator function of a lattice $\Lambda$ in the Euclidean space considering the squared length of $\mathbf{x} \in \Lambda$, defined as
\begin{align} \label{equ_theta_series}
    \theta= 1 q^0+ \sum_{i= 1}^{\infty} \tau_{d_i^2} q^{d_i^2},
\end{align}
where $\tau_{d_i^2}$ is the number of $\mathbf{x}$ having $\|\mathbf{x}\|^2 = d_i^2$ and $q$ is regarded as a dummy variable. 

\subsection{Construction A lattices}
Construction A is a method to form a lattice using a $q$-ary code $\mathcal{C}$ and an integer lattice $q\mathbb{Z}^n$. In this paper, we consider an important special case which uses binary codes as component codes, also known as the modulo-2 lattice.
\begin{myDef} \label{def_Cons_A}
    \rm (Modulo-2 construction A lattice \cite{zamir2014lattice}) Given an $(n, k)$ linear block code $\mathcal{C} \in \mathbb{F}_2^n$, a modulo-$2$ lattice is:
    \begin{align} \label{equ_Cons_A}
        \Lambda_a= \mathcal{C}+ 2 \mathbb{Z}^n.
    \end{align}
\end{myDef}
Let component code $\mathcal{C}$ have an $n \times k$ systematic generator matrix $\mathbf{G}_c= \begin{bmatrix}
    \mathbf{I}_k \\
    \mathbf{P}
\end{bmatrix}$, where $\mathbf{I}_k$ is a $k$-dimensional identity matrix. A lower triangular generator matrix of $\Lambda_a$ can be obtained as 
\begin{align}
    \mathbf{G}= \begin{bmatrix}
        \mathbf{I}_k & \mathbf{0} \\
        \mathbf{P}   & 2\mathbf{I}_{n- k}
    \end{bmatrix},
\end{align}
from which the volume of $\Lambda_a$ is
\begin{align} \label{equ_V}
    V(\Lambda_a)= 2^{n- k}.
\end{align}
Due to the presence of the $2 \mathbb{Z}^n$ lattice, the squared minimum distance of a construction A lattice is given by 
\begin{align} \label{equ_dmin_ConA}
    d_{min}^2= \min(4, d_c),
\end{align}
where $d_c$ is the minimum Hamming distance of $\mathcal{C}$. We use square $d^2$ to denote the Euclidean norm and the non-square $d$ to denote the Hamming distance, to distinguish the distance in the real number field and the binary field in this paper.

\subsection{Encoding and decoding construction A lattices}
\begin{figure}[t]
    \centering
    \includegraphics[width=0.9\linewidth]{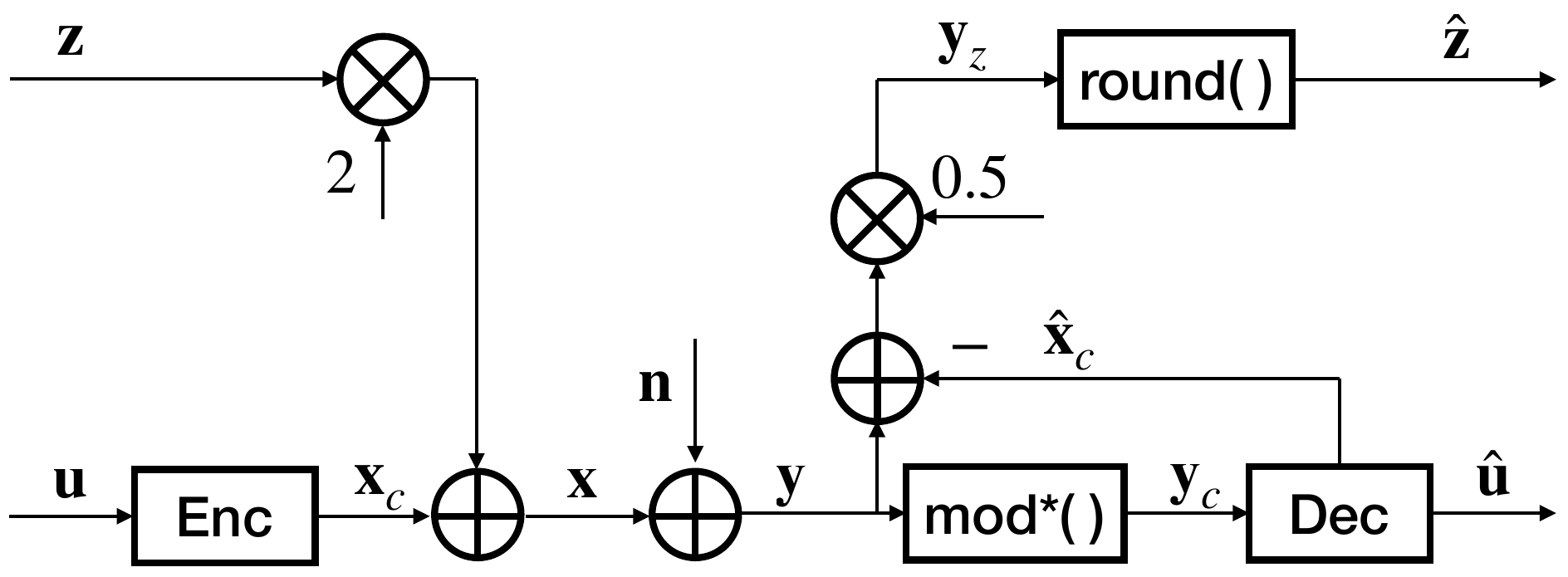}
    \caption{Encoder and decoder of construction A lattices.}
    \label{fig_ConsA_endec}
\end{figure}

Fig.~\ref{fig_ConsA_endec} gives the system model for the AWGN transmission using construction A lattices formed by an $(n, k, d_c)$ binary code $\mathcal{C}$. The module ``Enc'' and ``Dec'' are the binary encoder and decoder of $\mathcal{C}$. At the transmitter, a binary sequence $\mathbf{u} \in \{0, 1\}^k$ is first encoded to $\mathbf{x}_{\mathbf{c}} \in \mathcal{C}$ and then a lattice point is generated as $\mathbf{x}= \mathbf{x}_{\mathbf{c}}+ 2 \mathbf{z}$, where $\mathbf{z} \in \mathbb{Z}^n$. The AWGN channel output is
\begin{align}
    \mathbf{y}= \mathbf{x}+ \mathbf{n},
\end{align}
where Gaussian noise $\mathbf{n} \sim \mathcal{N}(0, \sigma^2 \mathbf{I})$.

At the receiver, successive cancellation decoding is applied for decoding construction A lattices. Before decoding $\mathcal{C}$, a modulo operation, introduced in \cite{conway1993sphere} and simplified in \cite{matsumine2018construction}, is performed to preserve distances to $(0, 1)$ as: 
\begin{align}
    \mathbf{y}_{\mathbf{c}}= \bmod^*(\mathbf{y})= |\bmod_2(\mathbf{y}+ 1)- 1|,
\end{align}
where operations are component-wise. The binary decoder finds an estimate of uncoded message $\hat{\mathbf{u}}$ and the corresponding coded message $\hat{\mathbf{x}}_{\mathbf{c}}$. Then subtract $\hat{\mathbf{x}}_{\mathbf{c}}$ from the received message to obtain $\mathbf{y}_{\mathbf{z}}= (\mathbf{y}- \hat{\mathbf{x}}_{\mathbf{c}})/ 2$, from which an estimate of integer $\hat{\mathbf{z}}$ is obtained by the rounding function.

\section{Lattice design with the truncated union bound} \label{sec_design_method}

The goal of this paper is to design a lattice to minimize the required VNR to achieve a given target WER, or equivalently minimize the WER for a given VNR. The theta series can be applied to compute the union bound of lattices. However, it has an infinite number of terms and requires the full weight enumerator function of $\mathcal{C}$. Instead, this paper considers the truncated union bound with first $m$ terms, where $d_m^2= d_c$, as:
\begin{align} \label{equ_UBE}
    P_e \approx \sum_{i= 1}^m \tau_{d_i^2} \cdot Q \left(\sqrt{\frac{d_i^2}{4 \sigma^2}} \right),
\end{align}
where $\tau_{d_i^2}$ and $d_i^2$ are from the theta series in \eqref{equ_theta_series}. To have a better estimate of $P_e$, we prefer to give the truncated theta series, denoted as $\theta'$, with as many terms as practicable. 
As a part of the theta series of $\Lambda_a$, the theta series of $2 \mathbb{Z}^n$ lattice can be obtained exactly by $\theta_{2\mathbb{Z}^n}= (\theta_{2\mathbb{Z}})^n$, where $\theta_{2\mathbb{Z}}$ is the theta series of the $2 \mathbb{Z}$ lattice as:
\begin{align}
    \theta_{2\mathbb{Z}}= 1 q^0+ 2 \sum_{i= 1}^{\infty} q^{4 i^2}.
\end{align}
Denote $\theta_{2\mathbb{Z}^n, d_i^2}$ as a truncation of $\theta_{2\mathbb{Z}^n}$ consisting of terms with $0< d^2 \leq d_i^2$, for example,
\begin{align*}
    \theta_{2\mathbb{Z}^4, 12}= 1 q^0+ 8 q^4+ 24 q^8+ 32 q^{12}.
\end{align*}

Let $\mathcal{C}$ be a binary code with known minimum Hamming distance $d_c$ and codeword multiplicity $\tau_c$, the number of codeword at $d_c$. A truncated theta series of the construction A lattice $\Lambda_a= \mathcal{C}+ 2 \mathbb{Z}^n$ can be explicitly written for terms having $d_i^2 \leq d_c$. For $d_c \leq 4$, the truncated theta series are
\begin{align} \label{equ_truncate_theta_1}
    \theta'= \begin{cases}
        1 q^0+ \tau_c 2^{d_c} q^{d_c} & d_c< 4 \\
        1 q^0+ (\tau_c 2^{d_c}+ 2n) q^{d_c} & d_c= 4.
    \end{cases}
\end{align}
For $d_c> 4$, let $L$ be the integer part of $d_c/4$. The truncated theta series are 
\begin{align}
    \theta'= \begin{cases}
        \theta_{2\mathbb{Z}^n, 4L}+ \tau_c 2^{d_c} q^{d_c}, & d_c \bmod 4 \neq 0 \\
        \theta_{2\mathbb{Z}^n, 4(L- 1)}+ \left(\tau_c 2^{d_c}+ \tau_{2\mathbb{Z}^n, d_c} \right) q^{d_c}, & d_c \bmod 4 = 0.
    \end{cases}
\end{align}

By \eqref{equ_VNR} and \eqref{equ_V}, 
the noise variance can be written as
\begin{align} \label{equ_sigma_R_VNR}
    \sigma^2= \frac{4^{1- R}}{2 \pi e \cdot \text{VNR}},
\end{align}
where $R= k/n$ is the code rate of $\mathcal{C}$. Since the truncated theta series $\theta'$ depends on the binary code $\mathcal{C}$, substituting \eqref{equ_sigma_R_VNR} to \eqref{equ_UBE}, the estimate of $P_e$ can be considered as a function of $\mathcal{C}$ and VNR as
\begin{align} \label{equ_VNR2Pe}
    P_e = f(\mathcal{C}, \text{VNR}) \approx \sum_{i= 1}^m \tau_{d_i^2} \cdot Q \left(\sqrt{\frac{d_i^2 \cdot 2 \pi e \cdot  \text{VNR}}{4 \cdot 4^{1- R}}} \right).
\end{align}
Since \eqref{equ_VNR2Pe} consists of multiple $Q$ functions, the inverse function is obtained numerically, denoted as
\begin{align} \label{equ_Pe2VNR}
    \text{VNR} = f^{-1}(\mathcal{C},\ P_e).
\end{align}
The best component code $\mathcal{C}_b$ is found by minimizing the required VNR for a target $P_e$:
\begin{align} \label{equ_design_criteria}
    \mathcal{C}_b= \mathop{\arg\min}_{\mathcal{C}} f^{-1}(\mathcal{C},\ P_e),
\end{align}
or equivalently, minimizing the estimate $P_e$ for a target VNR:
\begin{align}
    \mathcal{C}_b= \mathop{\arg\min}_{\mathcal{C}} f(\mathcal{C},\text{VNR}).
\end{align}
Equation \eqref{equ_design_criteria} is applied as the metric for giving design examples.

\section{Design examples using binary codes} \label{sec_design_exampla}

This section gives design examples based on the method in the previous section using EBCH codes and polar codes to form dimension $n= 128$ lattices. Since the truncated union bound estimate is accurate when $P_e$ is small and from a practical perspective, target WER for design examples are set for $P_e \leq 10^{-4}$. The best component EBCH codes and polar codes are given for different target $P_e$, from which the corresponding construction A lattices achieve the target $P_e$ with the lowest VNR, respectively. 

\subsection{Design using extended BCH codes}
\begin{figure}[t]
    \centering
    \includegraphics[width=0.8\linewidth]{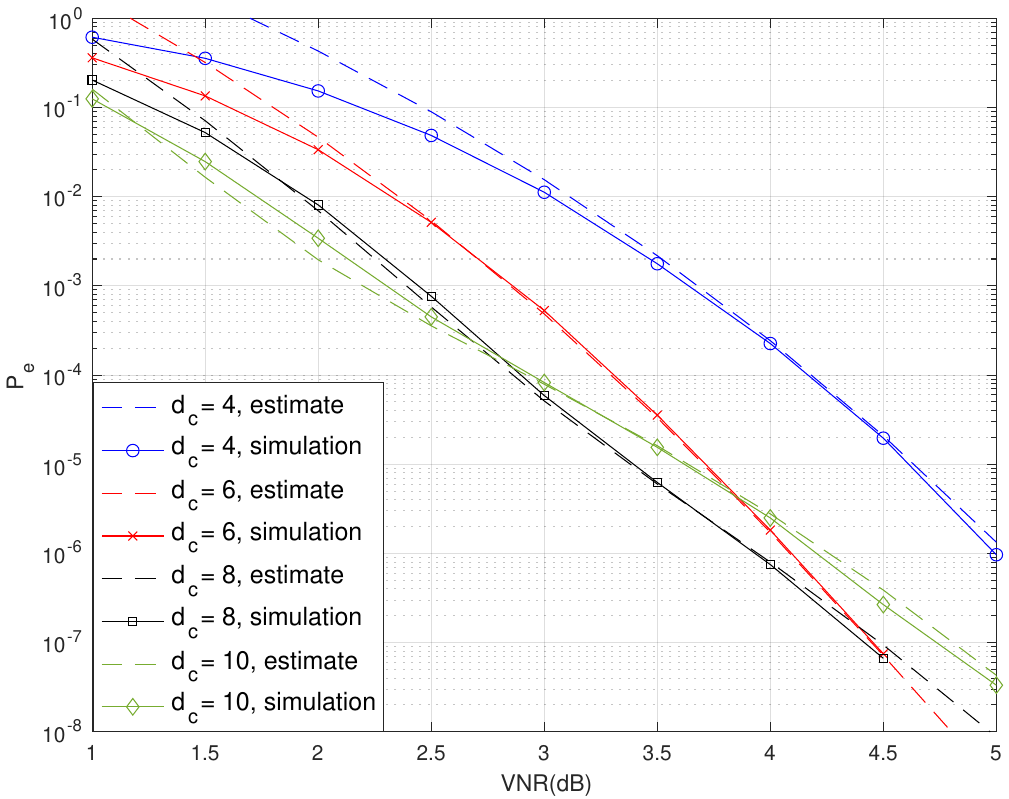}
    \vspace{-3mm}
    \caption{The truncated union bound estimate and numerical evaluation of $P_e$ for EBCH code lattice with $d_c= 4, 6, 8, 10$. Order-2 OSD algorithm is used to decode EBCH codes.}
    \label{fig_BCH_WER_UBE_dmin6810}
\end{figure}

The code parameters of $n= 128$ EBCH codes with given minimum Hamming distance $d_c$ can be easily obtained and the codeword multiplicity can be found in \cite{BCHweight2022} for computing the truncated union bound. 
Fig.~\ref{fig_BCH_WER_UBE_dmin6810} evaluates the truncated union bound in \eqref{equ_VNR2Pe}. The estimate of $P_e$ is verified by simulations using the order-2 OSD algorithm to decode EBCH codes. By \eqref{equ_design_criteria}, Table~\ref{table_BCH_ConsA} gives the best EBCH codes for different target $P_e$ from which the lattices achieve $P_e$ with the lowest VNR. 

The numerical results show that using $(128, 120, 4)$ EBCH code as component code, as suggested by the balanced distance rule, gives higher $P_e$ than other lattices using component codes with $d_c > 4$. This can be justified from the value of $\tau_4$, the number of lattice points at $d^2= 4$. 
Since $(128, 120, 4)$ EBCH code has $\tau_c= 85344$, then $\tau_4= 1365760$ by \eqref{equ_truncate_theta_1}; meanwhile for EBCH codes with $d_c> 4$, $\tau_4$ is only 256, from which smaller $P_e$ is achieved in Fig.~\ref{fig_BCH_WER_UBE_dmin6810}.
As VNR grows, the effect from high order terms in the truncated theta series $\theta'$ reduces on estimating $P_e$. At a relatively high VNR, estimate $P_e$ using lattice points with $d^2_{min}$ becomes sufficient and, for lattices with same $\tau_4$, the lattice with the smallest volume, equivalently the largest rate $R$ of the component code, gives the lowest $P_e$. 
This can be observed from Fig.~\ref{fig_BCH_WER_UBE_dmin6810}, that is the lattice using $(128, 113, 6)$ EBCH code achieves lower $P_e$ than that using $(128, 106, 8)$ EBCH code when $P_e\leq 10^{-7}$.
For very high VNR, since the $Q$ function in \eqref{equ_VNR2Pe} decreases exponentially as $R$ increases, the component code with highest $R$, that is the $(128, 120, 4)$ EBCH code here, will eventually become the best. 
The truncated union bound shows that this occurs at VNR $\approx 11.8$dB, where $P_e \approx 10^{-47}$, however this is beyond the scope of practical lattice design.

\begin{table}[t]
    \centering
    \caption{The code parameter of the best EBCH codes for construction A at given target $P_e$, with its required VNR to achieve $P_e$.}
    \begin{tabular}{|c|c|c|c|}
        \hline
        $P_e$     & Required VNR(dB) & Code parameter                   & $\tau_c$                \\ \hline
        $10^{-4}$ & 2.86             & \multirow{3}{*}{$(128, 106, 8)$} & \multirow{3}{*}{774192} \\ \cline{1-2}
        $10^{-5}$ & 3.38             &                                  &                         \\ \cline{1-2}
        $10^{-6}$ & 3.95             &                                  &                         \\ \hline
        $10^{-7}$ & 4.45             & \multirow{2}{*}{$(128, 113, 6)$} & \multirow{2}{*}{341376} \\ \cline{1-2}
        $10^{-8}$ & 4.81             &                                  &                         \\ \hline
    \end{tabular}
    \label{table_BCH_ConsA}
\end{table}

\subsection{Design using polar codes}
\begin{figure}[t]
    \centering
    \includegraphics[width=0.8\linewidth]{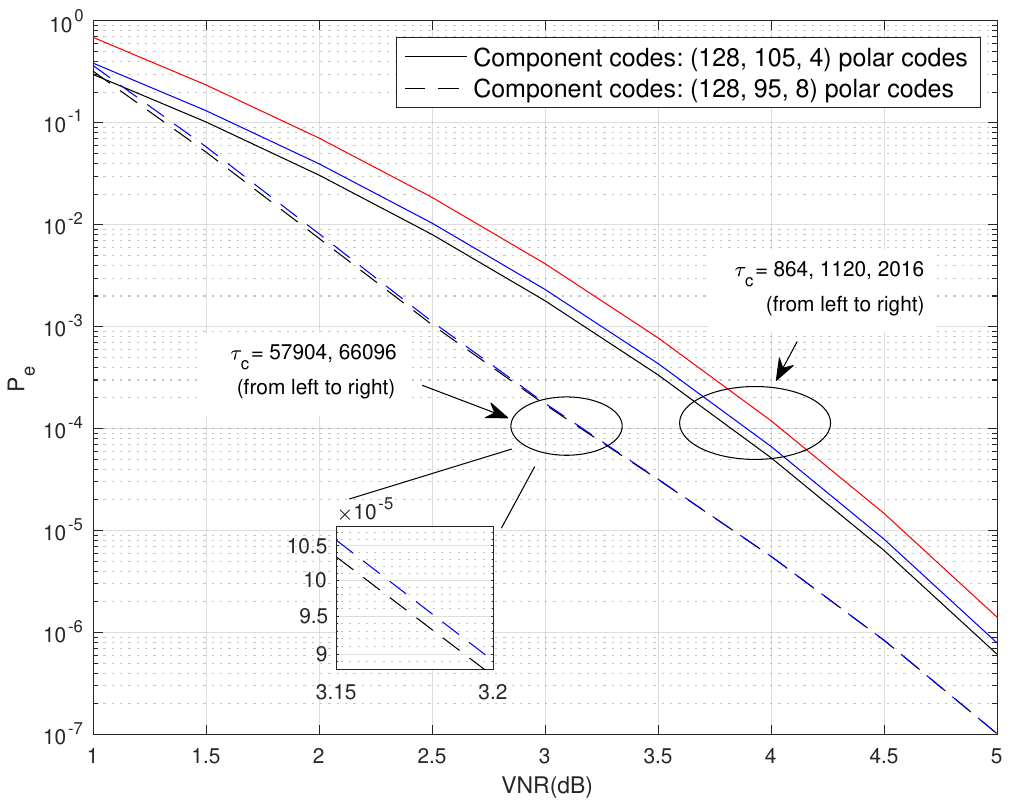}
    \vspace{-3mm}
    \caption{Truncated union bound of construction A lattices using polar codes with different $\tau_c$ for $d_c= 4, 8$.}
    \label{fig_polar_UBE_4_8_until_d_c_DMC}
\end{figure}

Polar codes are flexible in adjusting the code rate by steps as small as $1/n$. Furthermore, it is allowed to choose different information set to form polar codes with given code parameter $(n, k, d_c)$. By properly choosing the information set, $n= 128$ polar codes include $RM(r, 7)$ Reed-Muller codes, for $r= 0, \cdots, 7$, and the $(128, 120)$ extended Hamming code. 
To reduce the scope of the search for a good polar code, we first investigated a class of polar codes described in \cite{rowshan2023formation}, which are constructed based on the partial order of the binary representation of bit-channel indices. For such polar codes, $\tau_c$ can be computed analytically by the following proposition. 
Let $\mathcal{I} \subseteq [0, n- 1]$ be the information set of an $(n, k, d_c)$ polar code and a subset $\mathcal{I}' \subset \mathcal{I}$ indicate the rows of the polar transformation matrix with row weight $d_c$. 
\begin{myProposition} \label{prop_polar_tau_c}
    \rm \cite{rowshan2023formation} Consider an $(n, k, d_c)$ polar code satisfy the partial order property in \cite[Def. 2]{rowshan2023formation}. For all $i \in \mathcal{I}'$, let
    \begin{align*}
        \mathcal{K}_i= \Set{j \in [i+ 1, n- 1] | \rm{wt}(\mathbf{g}_j) \geq \rm{wt}(\mathbf{g}_i+ \mathbf{g}_j)= \rm{wt}(\mathbf{g}_i)},
    \end{align*}
    where $\rm{wt}(\mathbf{g}_i)$ is the weight of the $i$-th row of the polar transformation matrix. The codeword multiplicity is found by:
    \begin{align}
        \tau_c= \sum_{i \in \mathcal{I}'} 2^{|\mathcal{K}_i|}.
    \end{align}
\end{myProposition}
Proposition~\ref{prop_polar_tau_c} implies that $\tau_c$ only depends on the subset $\mathcal{I}'$. To design a polar code with a desired $d_c$ and rate $R$, all rows with weight larger than $d_c$ are first selected for $\mathcal{I}$ to avoid rate loss. Then select the set $\mathcal{I}'$ with row weight $d_c$ to form the full information set $\mathcal{I}$ to satisfy the partial order property given in \cite{rowshan2023formation} and the desired rate $R$. It is noticed that for a desired $d_c$ and rate $R$, there may exist multiple polar codes having different $\tau_c$.
This gives different truncated theta series $\theta'$ at the term with $d^2= d_c$. 
Due to the property of Gaussian distribution, changing $\tau_c$ for codes with larger $d_c$ gives less effect on estimating $P_e$ using \eqref{equ_VNR2Pe}.
This effect on truncated union bound estimate for $d_c= 4, 8$ is illustrated in Fig.~\ref{fig_polar_UBE_4_8_until_d_c_DMC}. When $d_c= 4$, choosing the component code with smaller $\tau_c$ improves $\approx 0.2$dB at $P_e= 10^{-4}$; however, when $d_c= 8$, such improvement is negligible.

\begin{figure}[t]
    \centering
    \includegraphics[width=0.85\linewidth]{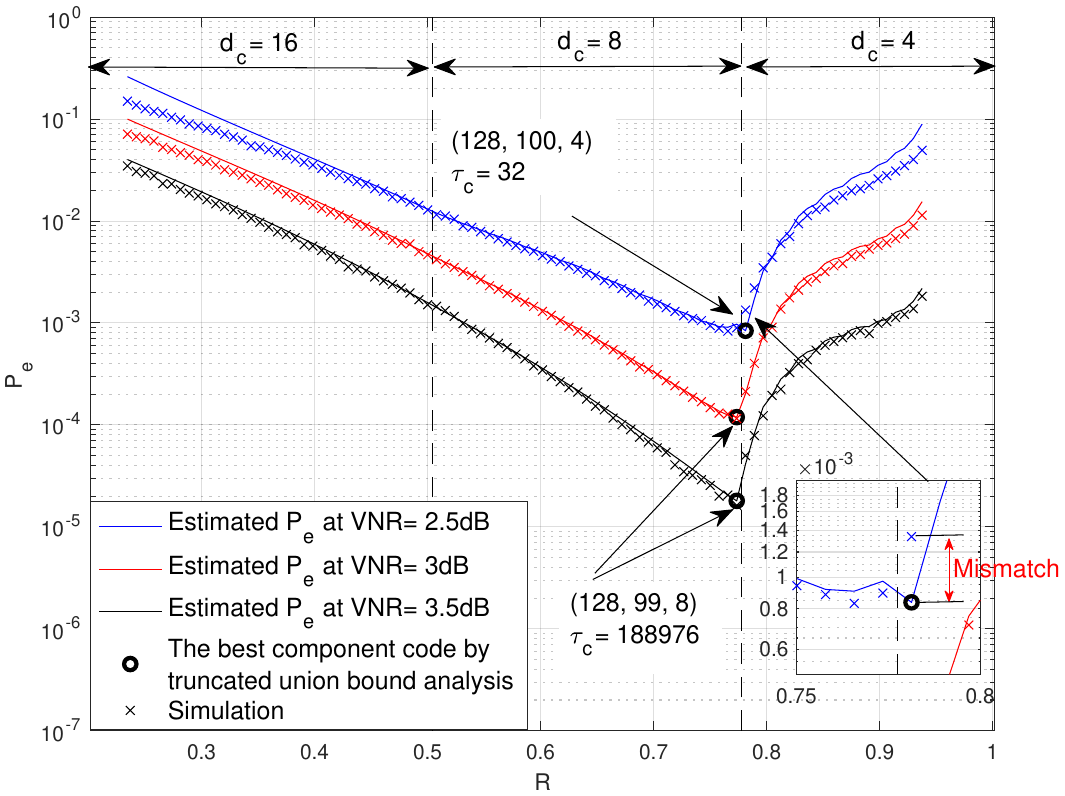}
    \vspace{-3mm}
    \caption{The truncated union bound estimate and numerical evaluation of $P_e$ for polar code lattice with different code rates for $d_c= 4, 8, 16$. Order-2 OSD algorithm is used to decode polar codes.}
    \label{fig_ConsA_polar_multitermUBE}
\end{figure}

Fig.~\ref{fig_ConsA_polar_multitermUBE} evaluates the truncated union bound estimate by \eqref{equ_VNR2Pe} of construction A lattices using polar codes with $d_c= 4, 8, 16$ as a function of rate $R$ at VNR $=2.5, 3, 3.5$dB. The polar codes satisfying the partial order property are considered. For polar codes with same $R$, the one with the lowest $\tau_c$ is selected as the candidate code for evaluation. The simulation results verify the truncated union bound estimation, where order-2 OSD is applied to decode polar codes.
At VNR$=2$dB with $P_e \approx 10^{-3}$, a mismatch of $P_e$ between the estimate and the simulation occurred, where the best component code has $d_c= 8$ from the simulation but not 4 from the estimate using \eqref{equ_VNR2Pe}. This is because the lattice points with $d^2>4$ may still contribute to $P_e$ at low VNR, which is not involved in \eqref{equ_VNR2Pe} for polar codes with $d_c= 4$. For error rate $P_e \leq 10^{-4}$, the contribution of lattice points with $d^2 > 4$ to $P_e$ becomes small and the estimate becomes accurate.
Even though the $(128, 99, 8)$ polar code has very large $\tau_c= 188976$, it reduces $\tau_4$ from 768, using the $(128, 100, 4)$ polar code, to 256 with only $1/128$ of rate loss, by which a lower error rate is achieved at target $P_e \leq 10^{-4}$. 
The best component polar code has code parameter $(128, 99, 8)$ and $\tau_c= 188976$, from which the lattice achieves $P_e= 10^{-4}$ to $10^{-8}$ at VNR $\approx 3.05, 3.67, 4.27, 4.82, 5.31$dB, respectively.

Additionally, as shown in Fig.~\ref{fig_polar_UBE_4_8_until_d_c_DMC}, lower $\tau_c$ improves $P_e$ of polar code lattice with $d_c= 4$. A search for polar codes not satisfying the partial order property is performed for rate $R= 100/128$ to $103/128$, which have performance close to the best in Fig.~\ref{fig_ConsA_polar_multitermUBE}. In this case, $\tau_c$ are found by numerical method. Lower $\tau_c$ are found at $R= 101/ 128, 102/128$ and $103/128$, however, better component polar code has not been found.

\section{Comparison with different design rules}  \label{sec_sim}
This section compares the WER performance for construction A lattices designed by the proposed truncated union bound analysis and by the classic balanced distance rule and the equal error probability rule, suggested in \cite{barnes1983new} and \cite{wachsmann1999multilevel}. For the balanced distance rule, the component EBCH code and polar code are selected with $d_c= 4$, giving the same Euclidean norm as $d_{min}^2$ of the $2 \mathbb{Z}^n$ lattice. For the equal error probability rule, the binary codes are selected to achieve same WER as the $2 \mathbb{Z}^n$ lattice across the additive modulo Gaussian noise (AMGN) channel \cite{ludwiniananda2021design}. Since EBCH codes only have fixed configurations, the equal error probability rule is applied only to polar code lattices. The order-2 OSD algorithm is used to decode binary codes of construction A lattices to provide near-ML performance, unless stated otherwise.

\begin{figure}[t]
    \centering
    \includegraphics[width=1\linewidth]{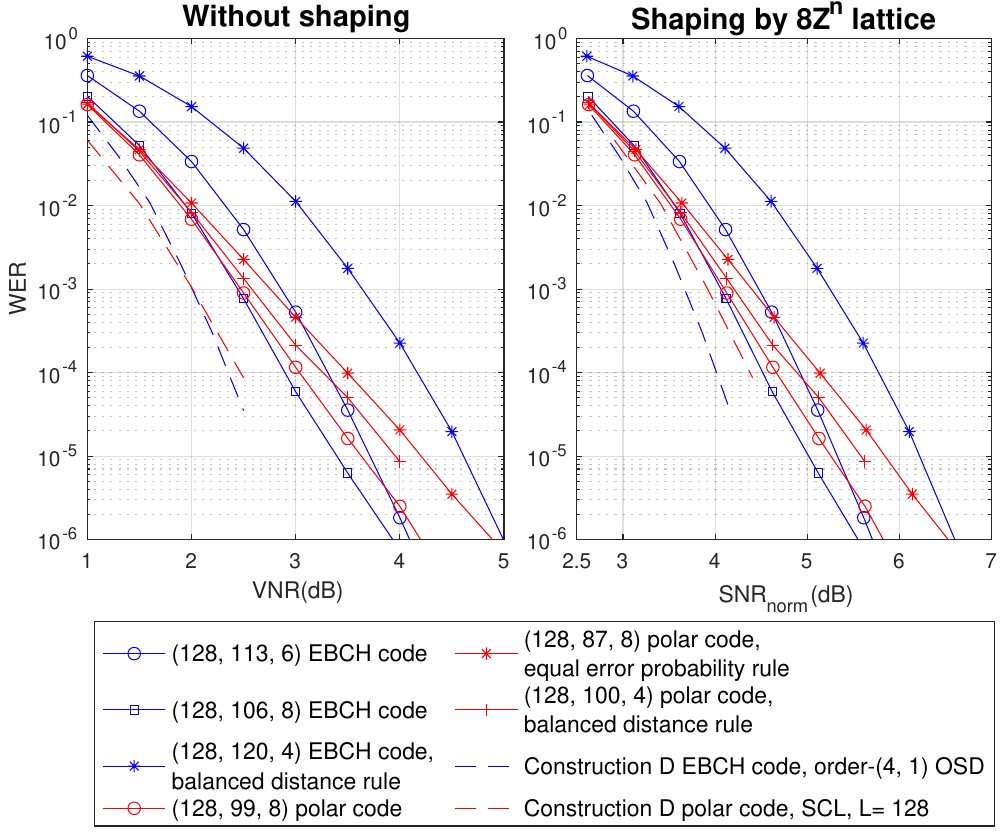}
    \vspace{-3mm}
    \caption{WER performance for construction A lattices with different design rules. Order-2 OSD is used to decode component codes of construction A lattices. Construction D lattices are from \cite{matsumine2018construction} and \cite{ludwiniananda2021design}.}
    \label{fig_ConsA_polar_BCH_diffrule}
\end{figure}

Fig.~\ref{fig_ConsA_polar_BCH_diffrule} shows WER of $n= 128$ lattices and lattice codes, shaping by $8 \mathbb{Z}^n$ lattice, using EBCH codes and polar codes, respectively. For lattice codes, SNR$_{norm}$ suggested in \cite{eyuboglu1993lattice} and \cite{tarokh1999universal} is considered, which allows one to compare lattice codes of different rates on the same scale, defined as
\begin{align}
    \text{SNR}_{norm}= \frac{P}{(2^{2 R_L}- 1) \cdot \sigma^2},
\end{align}
where $P$ and $R_L$ are the per-dimensional power and the code rate of the lattice code, respectively. By definition, $\text{VNR}$ and $\text{SNR}_{norm}$ are related as:
\begin{align}
    \text{SNR}_{norm}(dB) = & \text{VNR}(dB) + 10 \log_{10} (2 \pi e \cdot P) \nonumber \\
    & - \underbrace{10 \log_{10} [(2^{2R_L}- 1) \cdot V(\Lambda)^{2/n}]}_{\text{affected by code rate}}.
\end{align}
Among construction A lattices, the proposed design method achieves lower WER than that by the balanced distance rule and the equal error probability rule. At WER $=10^{-5}$, lattices using the $(128, 106, 8)$ EBCH code and the $(128, 99, 8)$ polar code are the best construction A lattices we found among those using EBCH codes and polar codes, respectively, and the $(128, 106, 8)$ EBCH code lattice is the best-known construction A lattice.
Construction D lattices in previous studies are also plotted for comparison using: EBCH code \cite{matsumine2018construction}, with component codes $\mathcal{C}_0= (128, 78, 16)$ and $\mathcal{C}_1= (128, 120, 4)$, using order-$(4, 1)$ OSD; polar code \cite{ludwiniananda2021design}, with $\mathcal{C}_0= (128, 7)$ and $\mathcal{C}_1= (128, 95)$, using successive cancellation list (SCL) decoding. Gaps remain from our best construction A $(128, 106, 8)$ EBCH code lattice to the construction D lattices. 
However, our design has lower decoding complexity and design cost. 
Our best lattice only needs $\sum_{i= 1}^2 {106 \choose i}= 5671$ OSD reprocessing computations in maximum for one decoding, compared to $\sum_{i= 1}^4 {78 \choose i}+ {120 \choose 1}= 1505702$ of the construction D EBCH code lattice. 
On the other hand, the proposed design gives an analytic approach to optimize the component codes using the truncated union bound; 
while the construction D polar code lattices with SCL applies the Monte-Carlo method, resulting in a higher design cost.

\section{Conclusion}
This paper designs construction A lattices using binary codes with known minimum Hamming distance and codeword multiplicity. The proposed design achieves lower WER than lattices designed using the classic rules for $n= 128$, where the $(128, 106, 8)$ EBCH code gives the best-known construction A lattice at WER of $10^{-5}$.
We can extend the design to higher dimensions for polar codes that satisfy the partial order property, where the capacity rule was applied in \cite{liu2018construction}. It is noticed that the best component polar code for $n= 128$ is also the $RM(4, 7)$ code. The study of higher dimensions may help to understand this relationship. 



\newpage
\footnotesize
	\bibliographystyle{ieeetr}
	\bibliography{Reference}

\end{document}